\documentclass[aps,prl,twocolumn,floats,showpacs,floatfix,a4paper]{revtex4}

\usepackage{epsfig}
\usepackage{color}
\usepackage{bm}
\usepackage{latexsym}

\begin{document}

\newcommand{\mm}[1]{{\mathbf{#1}}}
\newcommand{\cc}{{\bf\Large C }}
\newcommand{\hide}[1]{}
\newcommand{\tbox}[1]{\mbox{\tiny #1}}
\newcommand{\half}{\mbox{\small $\frac{1}{2}$}}
\newcommand{\sinc}{\mbox{sinc}}
\newcommand{\const}{\mbox{const}}
\newcommand{\tr}{\mbox{tr}}
\newcommand{\intt}{\int\!\!\!\!\int }
\newcommand{\ointt}{\int\!\!\!\!\int\!\!\!\!\!\circ\ }
\newcommand{\eexp}{\mbox{e}^}
\newcommand{\EPS} {\mbox{\LARGE $\epsilon$}}
\newcommand{\ar}{\mathsf r}
\newcommand{\im}{{\cal I}m}
\newcommand{\re}{{\cal R}e}
\newcommand{\bmsf}[1]{\bm{\mathsf{#1}}}
\newcommand{\dd}[1]{\:\mbox{d}#1}
\newcommand{\abs}[1]{\left|#1\right|}
\newcommand{\bra}[1]{\left\langle #1\right|}
\newcommand{\ket}[1]{\left|#1\right\rangle }
\newcommand{\mbf}[1]{{\mathbf #1}}
\newcommand{\eos}{\,.}
\definecolor{red}{rgb}{1,0.0,0.0}

\title{Experimental Observation of the dual behavior of ${\cal PT}$-symmetric scattering}

\author{Zin Lin, Joey Schindler, Fred M. Ellis, Tsampikos Kottos}
\affiliation{Department of Physics, Wesleyan University, Middletown, Connecticut 06459}

\date{\today}

\begin{abstract}
We investigate experimentally parity-time (${\cal PT}$) symmetric scattering using $LRC$ circuits in an inductively coupled 
${\cal PT}$- symmetric pair connected to transmission line leads. In the single-lead case, the ${\cal PT}$-symmetric 
circuit acts as a simple dual device - an amplifier or an absorber depending on the orientation of the lead. When a second 
lead is attached, the system exhibits unidirectional transparency for some characteristic frequencies. This non-reciprocal behavior is a consequence of 
generalized (non-unitary) conservation relations satisfied by the scattering matrix.
\end{abstract}

\maketitle

While there is absolutely no doubt as to the usefulness of gain mechanisms for signal boosting and information transfer,
loss on the other hand, is typically considered an evil - one to be avoided if at all possible - since it degrades the 
efficiency of the structures employed to perform useful operations on these signals. It is perhaps for this 
reason that researchers have never intentionally explored the combination of gain and loss as a duality of useful ingredients 
in device and materials engineering.

Currently however, an alternate viewpoint is emerging aiming to manipulate absorption, and via a judicious design that 
involves the combination of delicately balanced amplification and absorption mechanisms achieve classes of synthetic 
structures with altogether new physical behavior and novel functionality. This idea deliberately exploits notions of parity 
(${\cal P}$) and time (${\cal T}$) symmetry \cite{BB98,BBM99,BBJ02} and can be naturally incorporated into the framework of 
classical optics \cite{MGCM08}. In fact, optical media with delicately balanced gain and loss characteristic of systems with 
joint parity-time (${\cal PT}$) symmetry, have been reported \cite{RMGCSK10} showing several intriguing features \cite{MGCM08,
MMGC08,GSDMVASC09,RMGCSK10,FAHXLCFS11,RKGC10,ZCFK10,L09,L10d,L10b,CGS11,SXK10,CGCS10,M09,S10,LRKCC11}. These include among 
others, power oscillations and non-reciprocity of light propagation \cite{MGCM08,RMGCSK10,ZCFK10}, non-reciprocal Bloch 
oscillations \cite{L09}, and unidirectional invisibility \cite{LRKCC11}. In the nonlinear domain, such pseudo-Hermitian non
-reciprocal effects can be used to realize a new generation of on-chip isolators and circulators \cite{RKGC10}. Other results 
within the framework of ${\cal PT}$-optics include the realization of coherent perfect laser-absorber \cite{L10b,CGS11} and 
nonlinear switching structures \cite{SXK10}.

\begin{figure}
\includegraphics[scale=0.25]{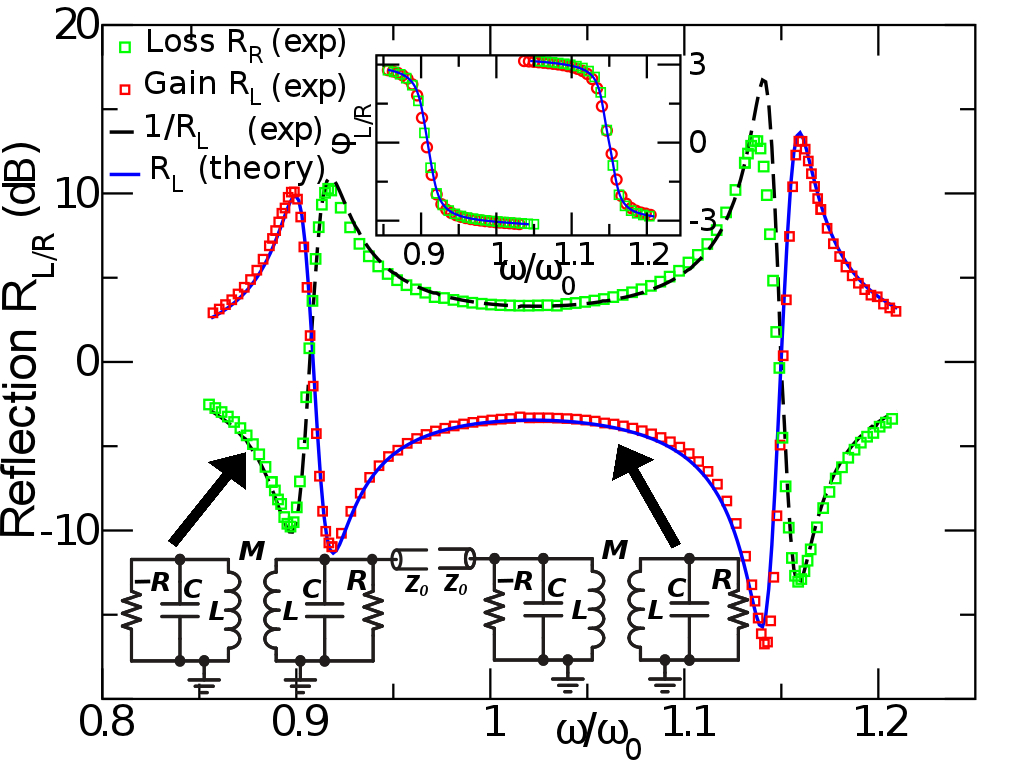}
  \caption{(Color) Experimental reflectances for a single TL attached to the lossy ($R_{\rm R}$) or the gain ($R_{\rm L}$) side 
of the dimer (see lower insets) for $\mu=0.29$, $\gamma=0.188875$, and $\eta = 0.0305$. The black line corresponds to $R_{\rm L}^{-1}$ 
and confirms the non-reciprocal nature $R_{\rm L} R _{\rm R}=1$ of the ${\cal PT}$-scattering. The upper inset shows the measurements 
for the left (right) reflection phases $\phi_L$ ($\phi_R$). The blue lines are the theoretical results Eq. (\ref{r_R}). 
}
\label{fig:fig1}
\end{figure}

More recently these ideas have been extended into the realm of electronic circuitry \cite{SLZEK11}, where it was demonstrated 
that a pair of coupled $LRC$ circuits, one with amplification and the other with equivalent amount of attenuation, provide the 
simplest experimental realization of a ${\cal PT}$ symmetric system. The ${\cal PT}$-circuitry approach suggested in Ref.~\cite{SLZEK11} 
opens new avenues for innovative integrated circuitry architectures which will afford novel avenues for signal manipulation, 
and reduced circuit loss. Moreover, it allow for direct contact with cutting edge technological problems appearing in (nano)-
antenna theory and split-ring resonator meta-material arrays. 

Although the study of ${\cal PT}$-symmetric Hamiltonians has been a subject of intense research efforts, relatively few authors 
have up to now theoretically studied the equivalent scattering system \cite{L10d,L10b,CGS11,LRKCC11,S10,CDV07,M09,CGCS10}. 
Given that the additional freedom of the gain/loss parameter will lead to a wealth of novel scattering phenomena, it is 
surprising that there is a complete lack of experimental investigations on ${\cal PT}$-scattering.

\begin{figure}
\includegraphics[scale=0.6]{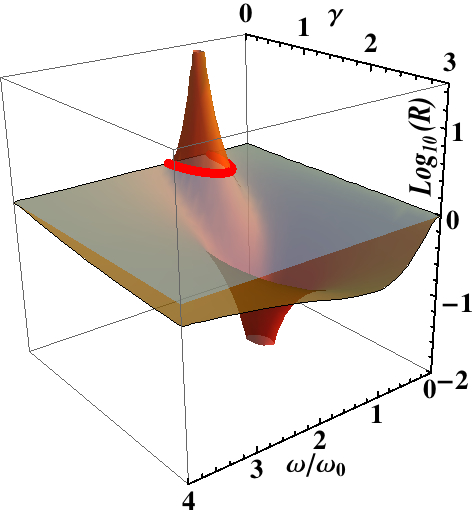}
  \caption{(Color) The $\omega-\gamma$ phase diagram for $\mu=0.57$, indicating the existence of a sub-unitary ($\log_{10} (R_R)
<0$) and a complimentary super-unitary ($\log_{10}(R_R)>0$) domain for the set-up shown at the lower left inset of Fig. \ref{fig:fig1}; 
The (while) plane $\log_{10} (R) = 0$ is shown for reference while the boundary Eq. (\ref{phase_t}) is indicated with a red line.}
\label{fig:fig2}
\end{figure}

Here we report initial experimental results for the scattering properties of ${\cal PT}$-symmetric systems. The scattering 
set-up consist of a pair of inductively coupled $LRC$ oscillators, one with gain and the other with loss, coupled to transmission
line (TL) leads. Our measurements reveal the signatures of the parity-time symmetry in the conservation relations satisfied by 
the non-unitary scattering matrix. In the simplest possible scattering set-up where the target is coupled to a single TL, we 
find that the reflection signal is non-reciprocal and respects the (non-unimodular) conservation relation $r_L\cdot r_R^*=1$ 
where $r_L (r_R)$ is the reflection associated to a left (right) incident wave. Furthermore, we have identified a transition 
from a sub-unitary to a super-unitary scattering process and associate it with the spatial structure of the potential inside 
the scattering domain. Once a second TL is attached to the  ${\cal PT}$-scatterer, the system demonstrates unidirectional 
transparency, where the transmittance is unity and the reflectance is zero, but only for waves incident from a single side.
Being free of basic theoretical approximations, and due to its relative simplicity in the experimental 
implementation, the $LRC$-networks with ${\cal PT}$ symmetry can offer new insights into the study of ${\cal PT}$-symmetric 
scattering which is at the forefront of current research in various areas of physics.

The heart of the ${\cal PT}$-symmetric scatterer (load) is the pair of inductively coupled $LC$ resonators (dimer) shown in 
the insets of Fig.~\ref{fig:fig1} \cite{SLZEK11}. Each inductor consists of 75 turns of \#24 copper wire wound on 15 cm diameter 
PVC forms in a $6\times6$ mm loose bundle for an inductance $L_0 = 2.32~mH$. The coils, matched to within 1\% by repositioning 
one of the turns, are mounted coaxially with a bundle separation that determines the mutual coupling of $\mu = M/L_0 = 0.29$ 
used for the data presented in this work. The capacitances are $10360~pF$ silver-mica in addition to the self capacitance of 
the coil bundles of $\sim 320~pF$. Capacitance balance is trimmed by substituting $\sim 360~pF$ of one side with a GR722-M 
variable capacitance. The uncoupled frequency of each resonator is $\omega_0=1/\sqrt{LC}=2\pi \nu$ with $\nu=30.957~kHz$. Loss 
imposed on the 
right half of the dimer is a standard carbon resistor, $R$. Gain imposed on the left half of the dimer, symbolized by $-R$, 
is implemented with an LF356-based negative impedance converter (NIC). The NIC gain is trimmed to oppositely match the value 
of $R$ used on the loss side, setting the gain/loss parameter $\gamma = R^{-1} \sqrt{L/C} = 1/(\omega_0RC)$. An additional NIC 
is included on the loss side so that intrinsic resonator losses on both sides can be compensated for prior to setting the 
gain/loss parameter.

We start our scattering studies with the following two reciprocal geometries: In the first case, a TL is attached to the left 
(amplified) circuit of the dimer load while in the second case, the TL is connected to the right (lossy) circuit of the load 
(see lower right and left insets of Fig. \ref{fig:fig1} respectively). Experimentally, the equivalent of a TL with characteristic 
impedance $Z_0$ is attached to either side of the dimer at the $LC$ circuit voltage node in the form of a resistance $R_0
=Z_0$ in series with an HP3325A synthesizer. The right and left traveling wave components associated with the TL are deduced 
from the complex voltages on both sides of $R_0$ with an EG\&G 7256 lock-in amplifier. With $V_{LC}$ the voltage on the $LC$ 
circuit, and $V_0$ the voltage on the synthesizer 
side of the coupling resistor $R_0$, the right (incoming) wave has a voltage amplitude $V_L^{+}=V_0/2$ and the left (reflected) 
wave has a voltage amplitude $V_L^{-}=V_{LC}-V_0/2$. The lock-in is referenced to the synthesizer, defining the phases of the 
wave components relative to the incoming wave.

At any point along a TL, the current and voltage determine the amplitudes of the right and left traveling wave components \cite{P05}. 
The forward $V_{L/R}^{+}$ and backward $V_{L/R}^{-}$ wave amplitudes, and $V_{L/R}$ and $I_{L/R}$ the voltage and 
current at the left (L) or right (R) TL-dimer contacts satisfy the continuity relation
\begin{equation}
\label{scatstates}
V_{L/R}=V_{L/R}^{+}+V_{L/R}^{-};\quad I_{L/R}=\left[V_{L/R}^{+}-V_{L/R}^{-}\right]/Z_0 
\end{equation}
which connect the wave components to the currents and voltages at the TL-dimer contact points. Note that with this convention, 
a positive lead current flows into the right circuit, but out of the left circuit, and that the reflection amplitudes for left or 
right incident waves are defined as $r_L\equiv {V_L^{-}/V_L^{+}}$ and $r_R\equiv {V_R^{+}/V_R^{-}}$ respectively.


Application of the first and second Kirchoff's laws at the TL leads allow us to find the corresponding wave amplitudes and reflection. 
For example, the case of the left-attached lead in the lower right inset of Fig. \ref{fig:fig1} gives
\begin{eqnarray}
\label{Kirchoff}
\eta (V_L^{+}-V_L^{-})= I_L^{M} - \gamma V_L - i\omega V_L \quad \quad  \nonumber \\
V_L = -i\omega \left[ I_L^M + \mu I_R^M\right]; \quad V_R = -i\omega \left[ I_R^M + \mu I_L^M\right] \\
0 = I_R^{M} + \gamma V_R - i\omega V_R \quad \quad \nonumber
\end{eqnarray}
where $\gamma$ is the gain/loss parameter, $\eta=\sqrt{L/C}/Z_0$ is the dimensionless TL impedance, and $I_{L/R}^M$ are the current 
amplitudes in the left or right inductors. Here, the dimensionless wave frequency $\omega$ is in units of $1/\sqrt{LC}$. Similar 
equations apply for the right-attached case shown in the lower left inset of Fig. \ref{fig:fig1}. 

We are interested in the behavior of the reflectance $R_{L/R}\equiv |r_{L/R}|^2$, and spacial profile of the potential 
$V_{L/R}$ inside the scattering domain, as the gain/loss parameter $\gamma$, and the frequency $\omega$ changes. 

\begin{figure}
\includegraphics[scale=0.3]{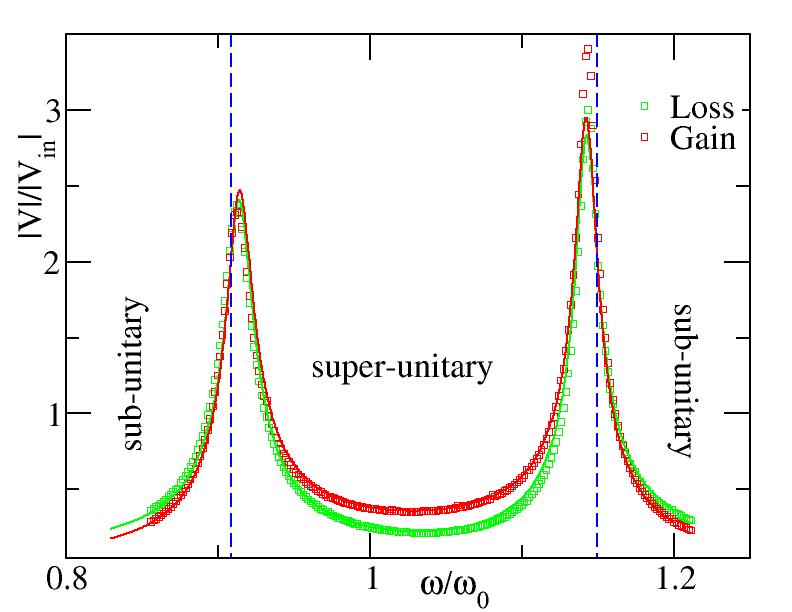}
  \caption{(Color) The spatial potential distribution inside the dimer versus the frequency $\omega$. 
The TL is coupled to the lossy side. We have used the same parameters
as those used in Fig. \ref{fig:fig1}. The blue dashed lines indicate the boundaries between sub-unitary
to super-unitary scattering where RD occur.
}
\label{fig:fig3}
\end{figure}

For ${\cal PT}$-symmetric structures, the corresponding scattering signals satisfy {\it generalized unitarity relations} 
which reveal the symmetries of the scattering target. Specifically, in the single-port set up this information is encoded 
solely in the reflection. To unveil it, we observe that the lower left set-up of Fig. \ref{fig:fig1} is the ${\cal PT}$-symmetric 
replica of the lower right one. Assuming therefore that a potential wave at the left lead (lower right inset) has the form 
$V_L(x)=\exp(ikx)+r_L\exp(-ikx)$ (we assume $V_L^+=1$ and $V_L^-=r_L$ in Eq. (\ref{scatstates})), we conclude that the 
form of the wave at the right lead associated with the lower left set-up of Fig. \ref{fig:fig1}, is $V_R(x)=\exp(-ikx)
+r_r\exp(ikx)= V_L^*(-x)$. Direct comparison leads to the relation
\begin{equation}
\label{r_Lr_R}
r_L\cdot r_R^* = 1 \rightarrow R_L= 1/R_R \quad {\rm and} \quad \phi_L=\phi_R
\end{equation}
where $\phi_{L/R}$ are the left/right reflection phases. Note that Eq. (\ref{r_Lr_R}) differs from the more familiar 
conservation relation $R=1$, which applies to unitary scattering processes as a result of flux conservation. In the 
latter case left and right reflectances are equal. Instead in the ${\cal PT}$-symmetric case we have in general that
$R_L\neq R_R$. 

For the specific case of the ${\cal PT}$-symmetric dimer, we can further calculate analytically the exact expression for 
the reflection coefficients. From Eqs. (\ref{Kirchoff}) we have
\begin{eqnarray}
\label{r_R}
r_L(\omega)=-f(-\eta, -\gamma)/f(\eta, -\gamma) \quad \quad \nonumber \\
r_R(\omega)=-f(-\eta, \gamma)/f(\eta, \gamma) \quad \quad \quad \\
f=1-\left[2-\gamma m(\gamma+\eta)\right]\omega^2+m\omega^4-i\eta \omega (1-m\omega^2) \nonumber \\
{\rm with} \quad m=1/\sqrt{1-\mu^2} \quad \quad \quad \nonumber
\end{eqnarray}
In the limiting case of $\omega\rightarrow 0,\infty$ the reflections becomes $r_R\rightarrow \mp 1$
and thus unitarity is restored.

In the main panel of Fig. \ref{fig:fig1} we report representative measurements of the reflection signals for the two
scattering configurations and compare them with Eq. (\ref{r_R}). 
The synthesizer frequency is 
slowly swept through the region of interest producing the reflectance $R_L=|V_L^{-}/V_L^{+}|$ as a function of frequency,
resulting in the red squares of Fig.~\ref{fig:fig1}.  A similar procedure is used to obtain the reflectance, $R_R$, 
from the right (loss) side of the dimer, resulting in the green squares of Fig.~\ref{fig:fig1}.
The measured reflectances $R_L$, and $R_R$ 
satisfy the generalized conservation relation $R_L\cdot R_R=1$ \footnote{The slight deviation from reciprocity in the 
vicinity of large reflectances in Fig.~\ref{fig:fig1} - see domain around the right peak - is attributed to nonlinear effects.} 
while for the reflection phases we have that $\phi_L=\phi_R$ as expected from Eq. (\ref{r_Lr_R}). 
Therefore, our experiment demonstrates that a ${\cal PT}$-symmetric load is a simple electronic dual device that for the same 
values of the parameters $\omega,\mu,\gamma$ acts as an absorber as well as a signal amplifier, 
depending on the direction of incident signal. 

Next, we identify the existence of a ${\it sub-unitary}$ domain for which $R<1$ (flux is deminished), and a ${\it super-unitary}$ 
domain for which $R>1$ (flux is enchanced). At the transition between the two domains $R_L=R_R=1$, in which case the scattering 
from both sides conserves flux.  Such {\it reflectance degeneracies} (RD) occur as a parameter such as the frequency $\omega$ (or 
$\gamma$) is varied continuously. Requiring that $|r_R|=1$, we get
\begin{equation}
\label{phase_t}
\gamma^*=\sqrt{-1+2\omega^2 -(1-\mu^2)\omega^4 \over (1-\mu^2)\omega^2}\quad {\rm and}\quad {|\omega^2-1|\over \omega^2}\leq \mu\leq 1
\end{equation}
A panorama of theoretical $R_R(\omega,\gamma)$ are shown in Fig. \ref{fig:fig2}. In the same plot we mark 
the transition line $\gamma^*(\omega)$ where a RD occurs. Inside this domain, a singularity point appears for 
which $R_R\rightarrow \infty$, while a reciprocal point for which $R_R=0$ is found in the complementary domain. 
The corresponding $(\omega_{\rm s};\gamma_{\infty,0})$ are found from Eq. (\ref{r_R}) to be
$\gamma_{\infty,0}={1\over 2} \left( \sqrt{\eta^2+{4\mu^2\over (1-\mu^2)}}\mp\eta\right);\, \omega_{\rm s}={1\over \sqrt{1-\mu^2}}$
Obviously via Eq. (\ref{r_Lr_R}) we have the reverse scenario for $R_L$. 

The sub-unitary to super-unitary ${\cal PT}$-symmetric transition is also manifest in the spatial structure of the potential 
$(V_L;V_R)$ inside the dimer. From Eq. (\ref{Kirchoff}) we get
\begin{eqnarray}
\label{V_LV_R}
V_L=2\eta \omega \left[1-m\omega (\omega-i\gamma)\right]/D \quad\quad\quad \nonumber \\
V_R=-2\eta \mu \omega /D\quad\quad\quad\quad\quad\quad \\
D=\eta\omega(1-m\omega^2)+i\left(1-\omega^2\left[2-m(\omega^2+\gamma\omega+\gamma^2)\right]\right)\nonumber
\end{eqnarray}
Typical potential amplitudes $(|V_L|;|V_R|)$, for the set-up of the lower left inset of Fig.  \ref{fig:fig1}, versus the frequency 
$\omega$ are shown in Fig. \ref{fig:fig3}. We observe that they are in general asymmetric. In the super-unitary domain, the gain 
side is characterized by a larger potential amplitude $|V_L|>|V_R|$ while in the sub-unitary domain the scenario is reversed and 
$|V_L|<|V_R|$. The latter configuration ensures that more power is being consumed than compensated by the gain circuit, while the 
inverse argument applies for the former configuration. At frequencies where the RD occurs, the potential 
profiles are spatially symmetric. This is consistent with the intuitive expectation that in order to conserve flux the excitation 
must on average spend equal amounts of time in the loss and gain circuits of the structure. Obviously, the reverse scenario occurs
if we coupled the ${\cal PT}$-dimer to the TL from the gain side.

\begin{figure}
\includegraphics[scale=0.3]{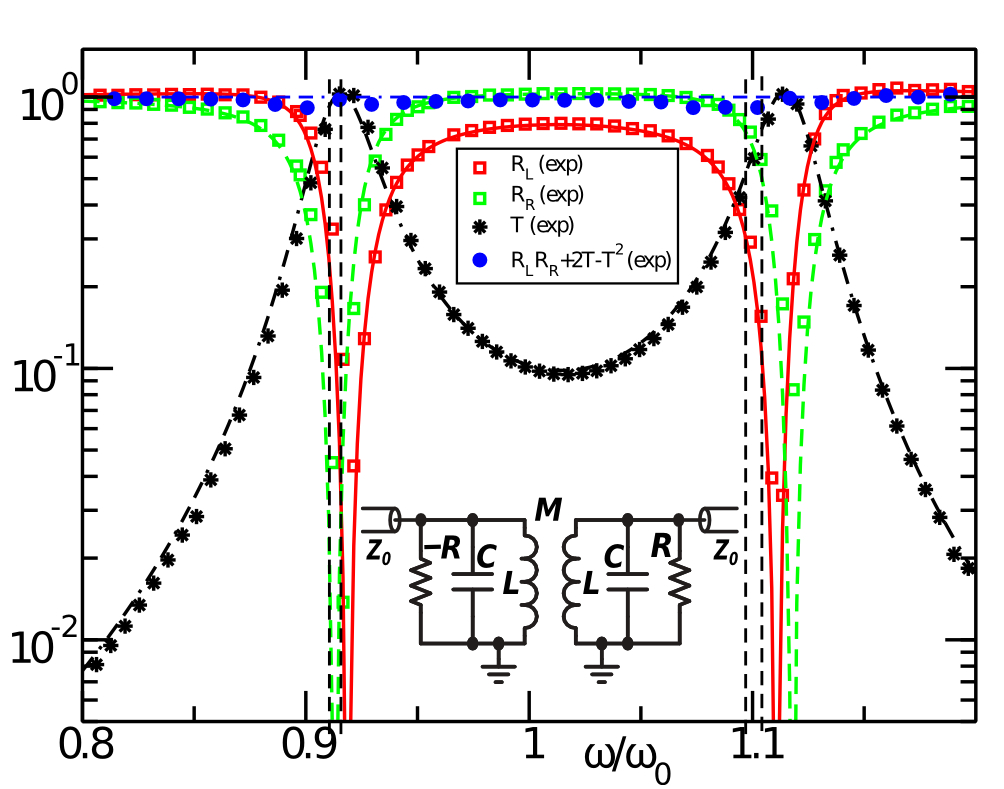}
  \caption{(Color) Experimental measurements (symbols) of $T$, $R_{L/R}$ for the two-port scattering set-up shown in the inset.
The solid lines (with the corresponding colors) are the numerical values of $T$, $R_{L/R}$. The conservation relation Eq. (\ref{TR_scat}) 
$R_L R_R + 2T -T^2=1$ extracted from the experimental data is reported with blue filled circles. The horizontal dashed blue line 
indicate the value $1$. The vertical dashed lines indicate the frequencies for which we have unidirectional transparency.
We have used the same parameters as those used in Fig. \ref{fig:fig1}. 
}
\label{fig:fig4}
\end{figure}

Finally, we have experimentally studied the generalized conservation relations for the case of two-port scattering processes 
(see inset of Fig \ref{fig:fig4}). Specifically, for 1D geometries, it was found \cite{L10b,CGS11,CDV07,LRKCC11} that while 
the reflectances for left and right incident waves might be different as in the single-port case, the corresponding 
transmittances are the same, i.e. $T_L=T_R=T$. Moreover, the following conservation relation was shown to hold \cite{CGS11}:
\begin{equation}
\label{TR_scat}
\sqrt{R_LR_R}=|T-1| 
\end{equation}
Note that Eq. (\ref{TR_scat}) is an intriguing generalization of the more familiar conservation relation $R+T=1$, which applies to 
unitary processes. In the ${\cal PT}$-symmetric case, the geometric mean of the two reflectances, $\sqrt{R_LR_R}$, replaces the 
single reflectance $R$ \footnote{Equation (\ref{r_Lr_R}) is a special case of Eq. (\ref{TR_scat}) once we realize that in the 
single port case the transmittance $T=0$}. 

Our measurements for $R_{L/R}$ and $T$ are shown in Fig. \ref{fig:fig4}. The quantity $R_L R_R + 2T -T^2$ (blue circles) is 
evaluated from the experimental data and it is found to be $1$ as it is expected from Eq. (\ref{TR_scat}). An interesting 
result of our analysis is that at specific $\omega-$values (marked with vertical dashed 
lines), the transmittance becomes $T=1$, while at the same time one of the reflectances vanishes. Hence, the scattering for 
this direction of incidence is flux conserving and the structure is {\it unidirectionally transparent}. Periodic 
repetition of the ${\cal PT}$-symmetric unit it will result in the creation of unidirectionally transparent frequency bands.
The phenomenon was first predicted in \cite{LRKCC11} and its generalization was discussed in \cite{CGS11}. 

In summary, we have presented experimental evidence of the anomalous properties of ${\cal PT}$-symmetric scattering. On 
this basis, we propose ${\cal PT}$-symmetric $LRC$ circuits as an easily realizable system where many other theoretical ideas 
can be investigated. Their simplicity and direct accessibility to the dynamical variables enables insight and a more thorough 
understanding of ${\cal PT}$-symmetric scattering. Due to space considerations, we defer discussion of other results pertaining 
to the time-dependent aspects of ${\cal PT}$-scattering to a subsequent publication.


\begin{acknowledgments}
We thank D. Christodoulides and V. Kovanis for useful discussions. Support from grants AFOSR No. FA9550-10-1-0433 
and NSF/ECCS-1128571 is acknowledged.
\end{acknowledgments}


\end{document}